\documentclass[fleqn,10pt]{wlscirepmod}
\pdfoutput=1
\usepackage{graphics}
 \usepackage{color}
\usepackage{wrapfig}
 \usepackage{bm}
 \usepackage{subcaption} 
\usepackage{ulem}
 \usepackage{lineno}
\usepackage{soul}
\usepackage{titlesec}
\usepackage{etoolbox}

\makeatletter
\patchcmd{\ttlh@hang}{\parindent\z@}{\parindent\z@\leavevmode}{}{}
\patchcmd{\ttlh@hang}{\noindent}{}{}{}
\makeatother

\newcommand\fbt{$f_{b}^{true}~$}

\newcommand\fbm{$f_{b}^{model}~$}
\newcommand{\Like}{\mathcal{L}}
\newcommand{\LL}{\mathrm{log}(\mathcal{L})}

\title {A model independent safeguard for unbinned Likelihood} 

\author[1,*]{Nadav Priel}
\author[2]{Ludwig Rauch}
\author[1]{Hagar Landsman}
\author[1]{Alessandro Manfredini}
\author[1]{Ranny Budnik}
 \affil[1]{Department of Particle Physics and Astrophysics, Weizmann Institute of Science, Rehovot, Israel}
 \affil[2]{Max-Planck-Institut f\"ur Kernphysik, Heidelberg, Germany}
 \affil[*]{nadav.priel@weizmann.ac.il}

\begin{document}
\begin{abstract}
We  present  a  universal  method  to  include residual un-modeled background shape uncertainties in likelihood based statistical tests for high energy physics and astroparticle physics.  
This approach provides a simple and natural protection  against mismodeling, thus lowering the chances of a false discovery or of an over constrained confidence interval, and allows a natural transition to unbinned space. Unbinned likelihood allows optimal usage of information for the data and the models, and enhances the sensitivity.

We show that the asymptotic behavior of the test statistic can be regained in cases where the model fails to describe the true background behavior, and  present 1D and 2D case studies for model-driven and data-driven background models. The resulting penalty on sensitivities follows the actual discrepancy between the data and the models, and is asymptotically reduced to zero with increasing knowledge. 
\end{abstract}
\maketitle
\section{Introduction}\label{sec:introduction}
One of the most challenging tasks when looking for a new phenomenon is the assessment of the background model and its uncertainties. 
In many cases the knowledge of the underlying background model is limited, and the model is estimated and tested against a  control measurement. In other cases, a reliable functional description of the background model exists, but  the relevant parameters fail to cover the true model.
This can happen, for example, when the model is based on Monte-Carlo (MC)  simulation with overly constrained  parameters, or with too many parameters, lacking the ability to track the effect of each of their combinations.

When using likelihood based statistical tests, it is assumed that the model is sufficiently flexible such that there exists a set of values of the parameters which can be regarded as ``true''. However, if the model does not reflect the truth accurately enough it might lead to an enhanced false discovery rate, or to an artificially enhanced exclusion limit.
  
There are several realistic cases in which it is challenging to provide a complete description of the background:
(i) The background model depends on ``too many'' parameters and it becomes computationally challenging to perform the minimization. (ii) The relevant parameters are hidden deep inside the MC unreachable by the user 
(iii) The shape is modeled based on a special dataset (i.e., calibration runs) with no underlying physical model that reliably accounts for the systematic uncertainties. 
(iv) The models, parameters or uncertainties used do not cover the true model. 

The excess events previously observed in some dark matter experiments \cite{Aalseth:2012if,Angloher:2011uu} have recently been
demonstrated to be entirely attributable to underestimated backgrounds \cite{Angloher:2014myn,Davis:2015vla,Davis:2014bla}, adding once more the importance of proper treatment of background uncertainties in these experiments and others.

A widely used procedure to establish discovery or exclusion in particle physics is based on a frequentist significance test using the Profile Likelihood (PL) ratio as the test statistic~\cite{Rolke:2004mj}.
The PL method has been proven useful to incorporate model uncertainties into the likelihood test.
 In addition to the parameter of interest (e.g., the cross section or number of signal events), the signal and background models contain nuisance parameters whose values are not taken as known a-priori. 
The additional flexibility introduced to parameterize systematic effects through the nuisance parameters results in a loss in sensitivity, according to the uncertainty on the parameters and their influence on the interpretation of the data.

In this paper we propose a procedure that provides a natural protection against mismodeling of the background, we use the unbinned PL ratio as a test statistic, and investigate its behavior. 

This paper is organized as follows: In Sec.~\ref{sec:likelihood} we briefly discuss the PL method and motivate the use of unbinned likelihood over the binned likelihood. In Sec.~\ref{sec:theproblem} we illustrate how mismodeling of the background causes loss of asymptoticness that enhances the false discovery rate (in case of underestimated background) or produces overly constrained exclusion limits (in case of overestimated background). In Sec.~\ref{sec:thesolution} we present a universal method to recover both the discovery potential and the exclusion sensitivity. In Sec.~\ref{sec:examples} we present three realistic use cases, with user-defined parametric background models and empirical non-parametric models using Kernel Density Estimates (KDE).

\section{Profile likelihood ratio test}\label{sec:likelihood}
\subsection{Likelihood function in counting experiments} \label{subsec:likelihood_function}
In high energy physics and astroparticle physics, we wish to identify a small signal over a non-negligible background.  In order to achieve that, the expected signal and background signatures in a detector have to show different behavior in some observable {\it x}. This  is modeled by the probability density functions (PDFs), $f_s(x)=f_s(x;{\bf \theta_{s}})$ for signal and $f_b(x)=f_b(x;{\bf \theta_{b}})$ for background, where $\theta_s$ and $\theta_b$ represent one or more model parameters.

If $N$ events were observed, the dataset is fully characterized be the vecotr  $\lbrace x_1,x_2,...,x_{N}\rbrace$ .
In a binned analysis, the observed space is divided to bins, so each bin contains $n_j$ observed events, and $\sum\limits_{j=1}^{N_{bins}} n_j = N$.

The probability for a signal (or background) event to be found in bin $j$ depends on the appropriate PDF, the bin location and its width
\begin{equation}
\label{eq:bins}
\epsilon_s^j= \int_{bin_j} f_s(x) dx  , \qquad \epsilon_b^j= \int_{bin_j} f_b(x) dx  \\,
\end{equation}

And the expected number of events per bin is
\begin{equation}
E[n_j] = N_s \epsilon_s^j + N_b \epsilon_b^j,
\end{equation}
where $N_s$ and $N_b$ are the total number of signal and background events expected over all bins.

The \textit{binned Likelihood} function is the product of the Poisson probabilities over all bins
\begin{equation}
 \label{eq:bLikelihood}
\Like_{b}=\prod_{j=1}^{N_{bins}}\mathrm{Poiss}(n_j | \epsilon_s^j N_s + \epsilon_b^j N_b).
\end{equation}

The binned scheme is simple, well defined and widely used, see e.g., \cite{Aprile:2011hx}. However, the projection of events into bins is associated with information loss. The choice of bins affects the sensitivity and the optimal choice varies for different hypotheses, detector performance and sample size (see chapter 11 in ~\cite{James:2006zz}).  

An alternative is the unbinned extended likelihood function (e.g., \cite{Barlow:1990vc})
\begin{equation}
 \label{eq:ubLikelihood}
\Like_{ub}=\mathrm{Poiss}(N | N_s + N_b) \prod_{i=1}^{N} \frac{N_s f_s(x_i) + N_b f_b(x_i)}{N_s + N_b}.
\end{equation}

Eq.~\ref{eq:ubLikelihood} is the infinitesimal version of Eq.~\ref{eq:bLikelihood}, and therefore contains all its information, and avoids the need of optimizing bins in order to enhance the sensitivity, as we show  in appendix~\ref{appendix:Abinned}. The unbinned scheme is invariant under coordinate transformation, as demonstrated in  appendix~\ref{appendix:Atx}.

Throughout this paper we will work with the unbinned likelihood function described above, with  $N_s$ as the \textit{parameter of interest}, and $N_s=0$ corresponding to the null, background only hypothesis.
$N_b$ is treated as a nuisance parameter. Though not explicitly shown here,  additional nuisance parameters and their constraint terms can be added to the likelihood function in the usual way.

\subsection{Profile Likelihood test for new physics}
\label{subsec:likelihood_PL}

The PL statistical inference approach is  widely used  for quantifying the level of agreement between a dataset and a null or signal hypothesis and its asymptotic formulae have been well studied~\cite{Cowan:2010js}.

In this approach, given a likelihood function  $\Like$ which is a function of a \textit{parameter of interest} ( $N_{s}$ in our case)  and a set of  \textit{nuisance parameters}, $\bm{\theta}_i$, the following test statistic is constructed
\begin{equation}
\label{eq:testStat}
q_{\mu} = -2 \mathrm{log} \Bigg( \frac{\Like(N_s,\hat{\hat{\bm{\theta_i}}})}{\Like(\hat{N_s},\hat{\bm{\theta_i}})} \Bigg).
\end{equation}

$\hat{\hat{\bm{\theta}}}_i$ are the values of the nuisance parameters that maximize $\Like$ for a given  $N_s$, and $\hat{N_s}$ and $\hat{\theta_i}$ are the maximum likelihood estimators (MLE) of the maximized (unconditional) likelihood function.

If for a specific dataset, $q_{N_s,obs}$ was observed when testing a specific $N_s$,  the level of disagreement between the data and the $N_s$ hypothesis  can be quantified using the $p$-value or the equivalent significance, $Z$,
\begin{equation}
\label{eq:pValue}
p_{N_s} = \int_{q_{N_s,obs}}^{\infty} f(q_{N_s}|N_s) dq_{N_s} , \qquad  Z=\Phi^{-1}(1-p),
\end{equation}

where $f(q_{N_s}|N_s)$ is the distribution of $q_{N_s}$ under the assumption of signal strength $N_s$, and $\Phi^{-1}$ is the quantile of the standard Gaussian. 
According to Wilks~\cite{Wilks:1938dza}, if $\hat{N_s}$ has a Gaussian distribution around its true value, and certain regularity conditions are met, $f(q_{N_s}|N_s)$ asymptotically  approaches a $\chi^2$ distribution with one degree of freedom \footnote{Throughout this paper we assume a single parameter of interest}.

For the statistical inference of discovery claim or exclusion limits we follow the procedure described in~\cite{Cowan:2010js}:

For a discovery claim the null hypothesis $N_s=0$ is to be rejected only when the data deviates from it  with a positive signal $\hat{N_s}>0$. Cases  in which $\hat{N_s}<0$ are not considered as an evidence against the null hypothesis

\begin{equation}
 \label{eq:discoveryTestStat}
q_0=\left\{
\begin{array}{ll}
-2 \mathrm{log} \Big( \frac{\Like(0;\hat{\hat{\theta}})}{\Like(\hat{N_s},\hat{\theta})} \Big), & \hat{N_s} \geq 0\\
      0, & \hat{N_s}<0.
\end{array}
\right.   
\end{equation}
and $f(q_{0}|0)$ can be asymptotically described by half a $\chi^2$ distribution plus half a delta function at zero. 

\begin{wrapfigure}{r}{0.5\textwidth}
  \begin{center}
    \includegraphics[width=0.48\textwidth]{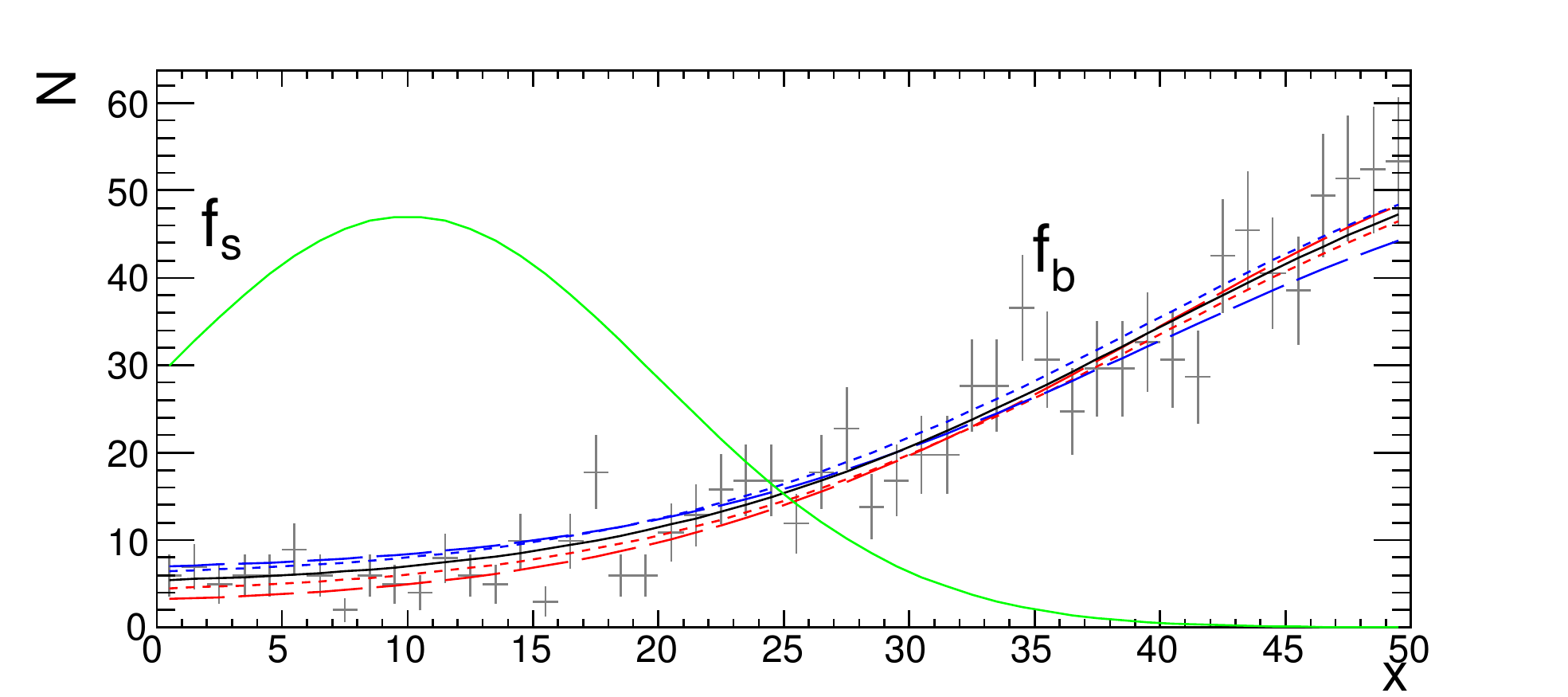}
  \end{center}
  \caption{ The signal model $f_s$ (in green, left), the true background model \fbt in black overlaid on a 1000 events background only dataset (generated using \fbt). The two dashed lines above 
it on the left side ($x<25$) represent \fbm with an overestimation in the signal region, and the two below it represent underestimation of the background in the signal region. }
\label{fig:fb_models}
\end{wrapfigure}

For an exclusion limit we  aim at rejecting a signal hypothesis $N_s$, and do not consider data with $\hat{Ns}>N_s$ as being less compatible with the $N_s$ hypothesis

\begin{equation}
 \label{eq:exclusionTestStat}
q_{N_s}=\left\{
\begin{array}{ll}
-2 \mathrm{log} \Big( \frac{\Like(N_s;\hat{\hat{\theta}})}{\Like(\hat{N_s},\hat{\theta})} \Big),  & \hat{N_s} \leq N_s\\
0, & \hat{N_s}>N_s,
\end{array}
\right.   
\end{equation}

and $f(q_{N_s}|N_s)$  can be asymptotically described by half a $\chi^2$ distribution plus half a delta function at zero.

\section{The problem}
\label{sec:theproblem}

Improper modeling of  background can cause devastating consequences such as a false discovery claim or  overly constraint limit. To illustrate this, toy datasets were generated using a ``true''  background model \fbt, and unbinned analysis was performed as in Eq.~\ref{eq:ubLikelihood} alas using a ``slightly different'' background model \fbm. Fig.~\ref{fig:fb_models} shows  \fbt and its four variations. Two of the variations  overestimate the background in the signal region (which can cause  overly constraint exclusion limits), and two underestimate it (which can lead to false discovery claims). While this modeling scenario is somewhat artificial, it aims at illustrating the challenges   in a coherent way. More realistic examples will be presented in Sec.~\ref{sec:examples}.

\begin{figure}[h] {\includegraphics[width=0.5\textwidth]{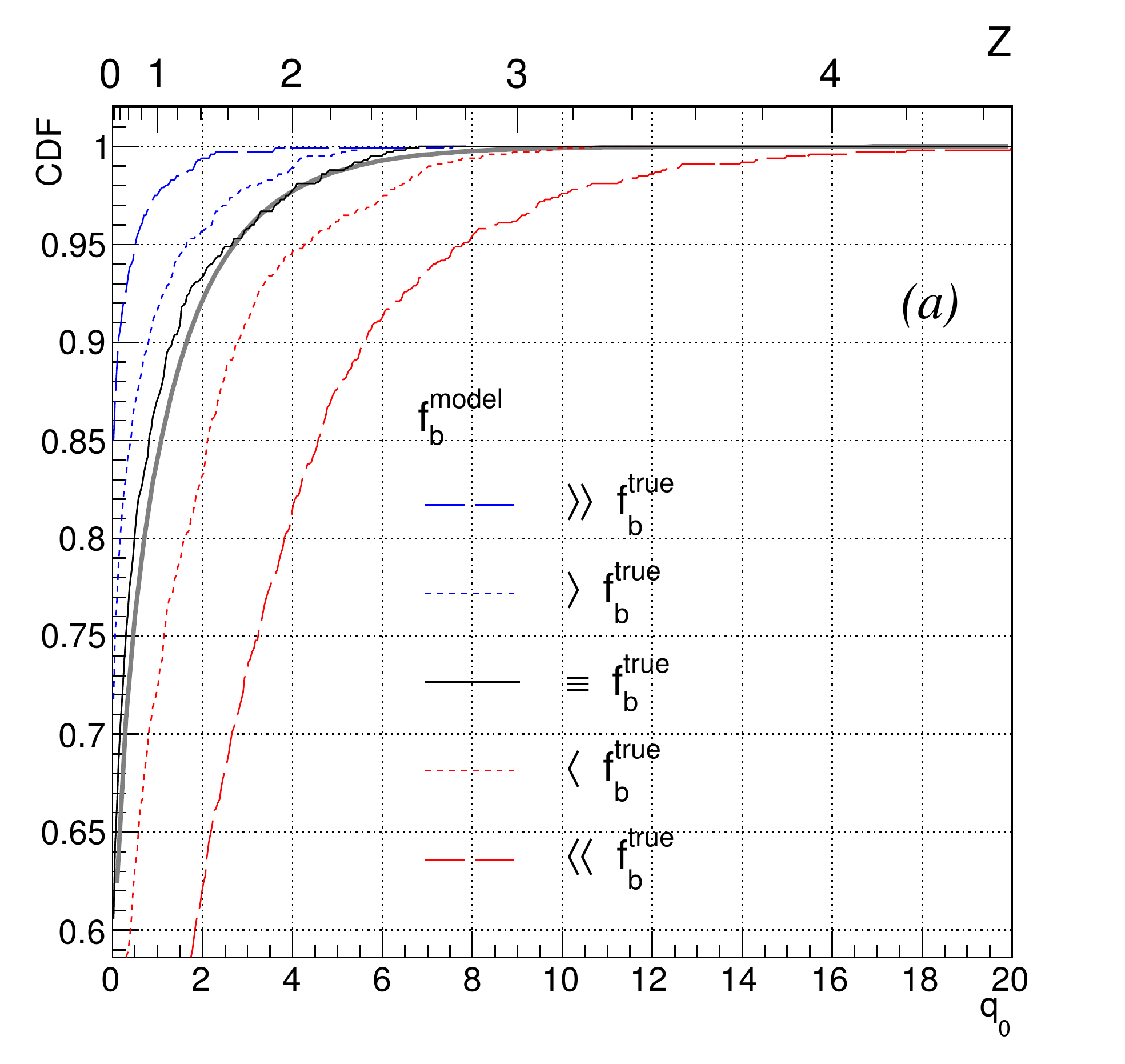}} 
{\includegraphics[width=0.5\textwidth]{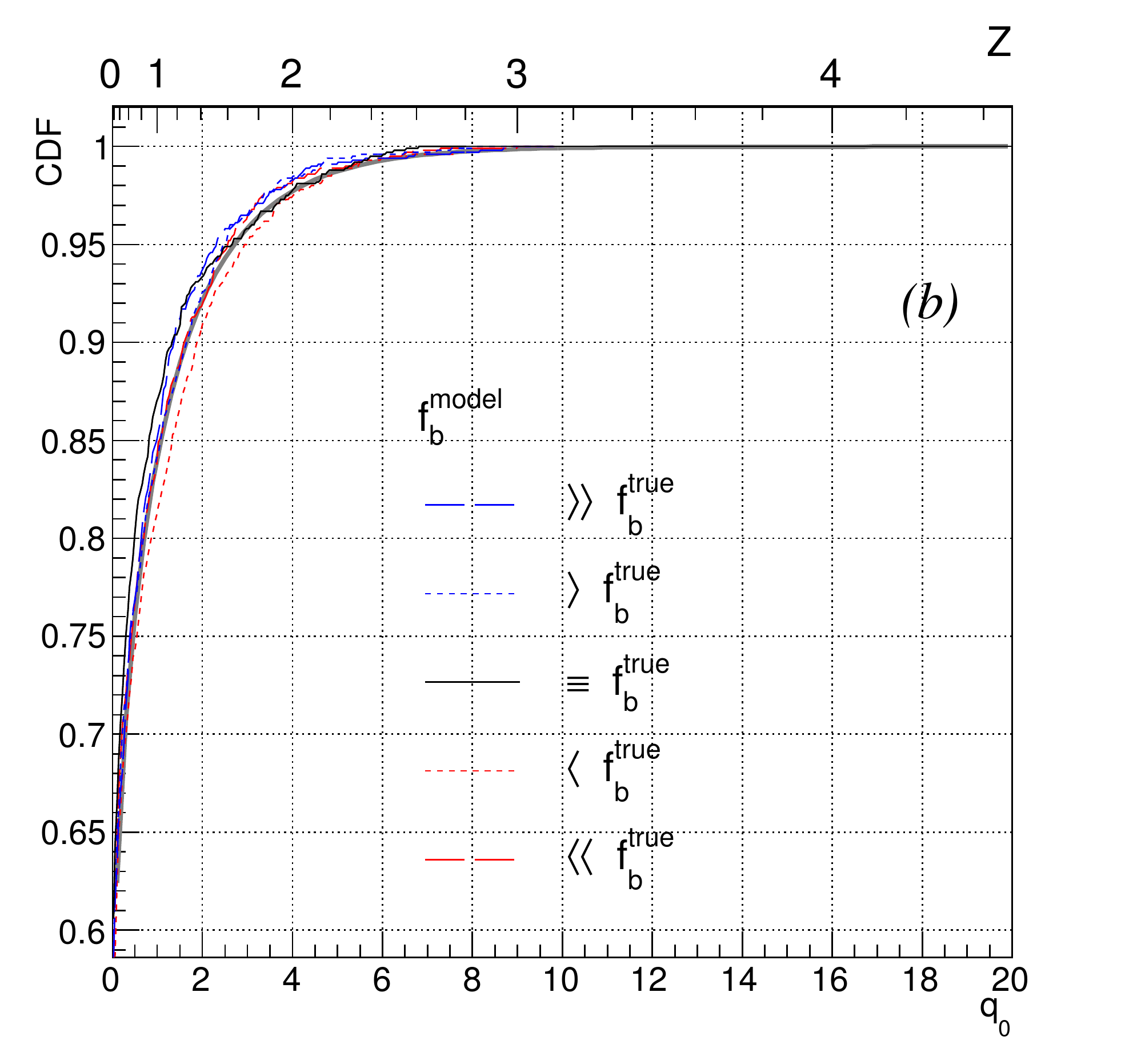}}
\caption{ The CDF of the discovery test statistic, $f(q_0|0)$ using the true background model (solid) and 4 variations as shown in Fig.~\ref{fig:fb_models}. {\bf (a)} with the regular procedure showing under (over) coverage when using background models that underestimate (overestimate)  the background in the signal region. {\bf (b)} with the correction the asymptotic behavior is regained. The expected $\frac{1}{2}\chi^2$ asymptotic behavior is shown in gray.}
\label{fig:discovery_user}
\end{figure}

To test the false discovery rate Eq.~\ref{eq:discoveryTestStat} was used on background only datasets using the true background shape \fbm=\fbt , and each one of the four \fbm variations .  Fig.~\ref{fig:discovery_user}(a) shows for each \fbm the cumulative distribution function (CDF) of  $q_0$ ($f(q_0|0)$). The deviation from the assumed (cumulative) $\frac{1}{2}\chi^2$ distribution is evident. In case of underestimated background model the false discovery rate is enhanced, and the overestimated background model artificially strengthens the exclusion limit as will be discussed in more details along with the proposed solution in Sec.~\ref{sec:thesolution}.

In the next section we will show how to account for these unmodeled mismatches of the background, in a way that preserves the coordinate invariance and the optimal sensitivity.

\section{The solution}
\label{sec:thesolution}

In this section we present a Poka-yoke method to handle mismodeled or overlooked uncertainties in unbinned analyses, verify its asymptotic behavior, and discuss its implementation for discovery and exclusion.

Starting with a benchmark background model \fbm , and a background only calibration measurement we quantify the level of ``signal likeness'' that can be overlooked in the calibration sample by introducing a \textit{contamination parameter}, $\epsilon$.  The benchmark background model will be added  a signal like component quantified by $\epsilon$ which will serve as a nuisance parameter,

\begin{equation}
\label{eq:backgroundVariation}
f_b^{model}(x) \rightarrow (1-\epsilon)f_b^{model}(x) + \epsilon f_s(x).
\end{equation}

This variation is motivated in appendix~\ref{appendix:B} using two approaches. 
This variation is universal and it does not depend on the way the background is modeled. In addition, it is unique with respect to other functional variations of the background, as it is the most conservative one with respect to the tested hypothesis.

As a result of this extra variation the likelihood function (i.e.,  Eq.~\ref{eq:ubLikelihood}) is promoted to 
\begin{equation}
 \label{eq:correctedLikelihood}
\Like_{phys}=\mathrm{Poiss}(N | N_s + N_b) \prod_{i=1}^{N} \frac{N_s f_s(x_i) + N_b (1-\epsilon) {f_b}(x_i) + N_b \epsilon f_s(x_i)}{N_s + N_b}.
\end{equation}
The signal-free calibration dataset provides additional constraints on $\epsilon$,
\begin{equation}
 \label{eq:EpsilonLikelihood}
\Like_{\epsilon} = \prod_{i=1}^{N_{cal}} (1-\epsilon) {f_b}(x_i) + \epsilon f_s(x_i),
\end{equation}
combined to
\begin{equation}
\label{eq:likelihood}
\Like (N_s; N_b,\epsilon)=\Like_{phys} \times \Like_{\epsilon}.
\end{equation}

Using this equation with Eqs.~\ref{eq:discoveryTestStat} and~\ref{eq:exclusionTestStat} this method can be used for discovery or exclusion tests. 

The method can be made more conservative by limiting $\epsilon$ to be only positive (for conservative discovery tests) or only negative (for conservative exclusion tests). However, as will be soon demonstrated this does not seem to be necessary, and the asymptotic behavior is recovered without any constraints on $\epsilon$.    

While summarizing this paper, we became aware that a similar approach is used by the ATLAS collaboration. In~\cite{Aad:2014eha}, the ``spurious signal'' method was used to systematically compare background models, in a similar way to the method presented here. However, in this paper, we show that it is  a valid and necessary addition to the likelihood in order to model shape uncertainties. In addition, we motivate the method (appendix~\ref{appendix:B}), and investigate its test statistic asymptotic behavior in a systematic way. 
We also demonstrate how it can be used in different types of analyses, in order to restore the assumed asymptotic behavior (Sec.~\ref{sec:examples}), both for exclusion and discovery of a new signal.

\subsection{Discovery}\label{subsec:userModel}
To test the asymptotic behavior we repeat the procedure described in Sec.~\ref{sec:theproblem} in conjunction with the procedure described here.
As can be seen in Fig.~\ref{fig:discovery_user} (b),  using the new procedure the test statistic distribution regains its expected asymptotic behavior,  for both over-~and underestimation of the background in the signal region. In both cases, the $\frac{1}{2}\chi^2$ is an excellent approximation and therefore the significance (or $p$-value) estimated using the PL will be reliable, and will not lead to false discovery. 

To see how this procedure affects the discovery sensitivity we present it using fake data samples with 15 signal events over 100 background events\footnote{Throughout this paper we always note the expected number of events, where for each trial this number is Poisson distributed.}. This is done using the true background model, and the 4 background model variations described in Fig.~\ref{fig:fb_models}. 

\begin{figure}[h!]
\centerline{\includegraphics[width=1.0\linewidth]{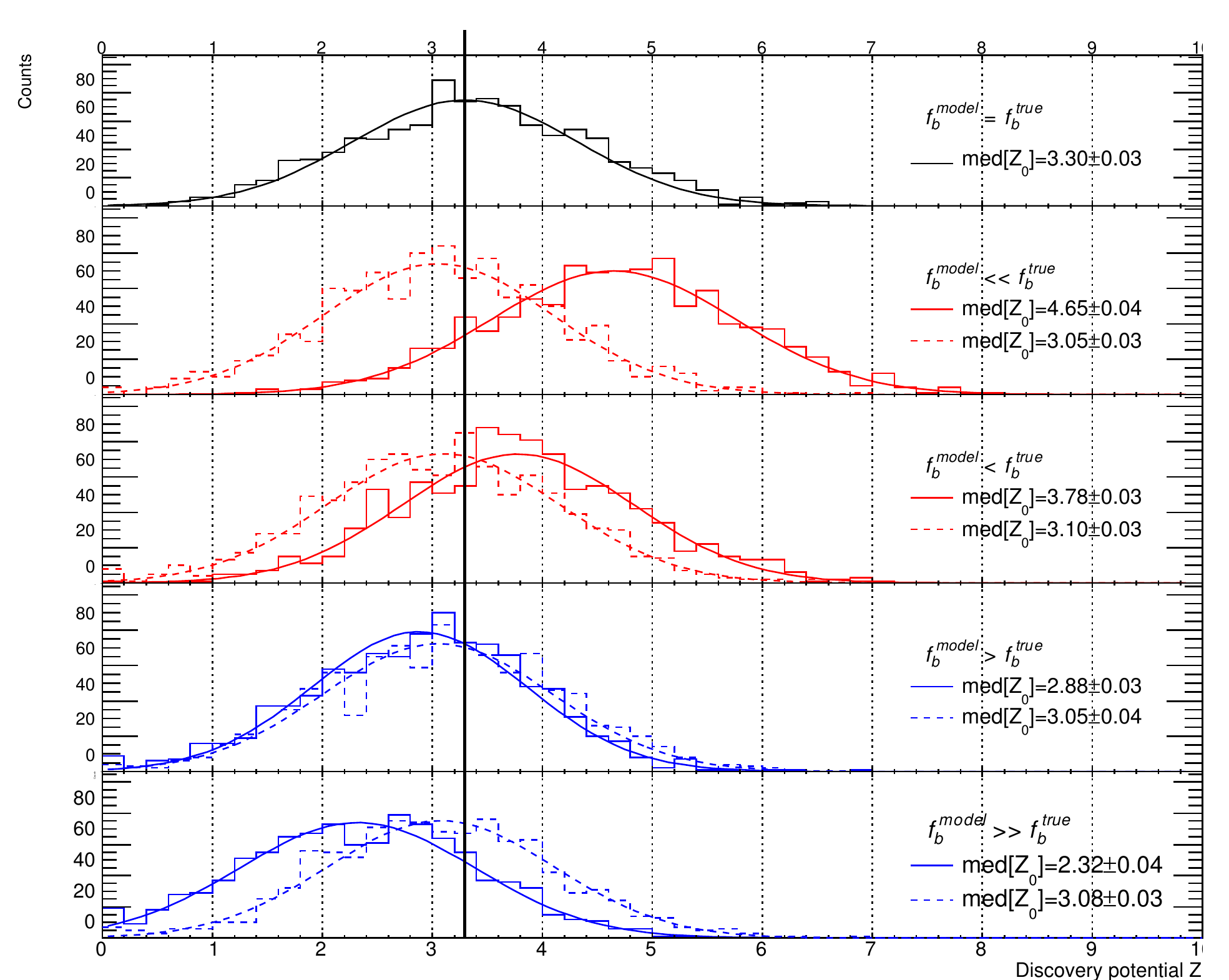}}
\caption{Discovery potential for 15 signal events over 100 background events. The sensitivity of the over-~or underestimated background models (solid lines) is compared with the sensitivity of the method described in this paper (dashed lines). The upper panel (black) shows the sensitivity using the true underlying model. The second and third panels show the sensitivity for big and small underestimations. The forth and fifth panels show the sensitivity using overestimated background models. In each and every case the sensitivity using the suggested method (dashed lines) is closer to the true sensitivity of the experiment (upper panel).}
\label{fig:discovery_significance}
\end{figure}

In Fig.~\ref{fig:discovery_significance} on the upper panel (in black) the sensitivity using the true background is shown. We would like to have a distribution as similar as possible to this. If a distribution is shifted to the right, it means that we get an artificial enhancement of the sensitivity and we are exposed to false discoveries. Alternatively, a left shifted distribution with respect the true one signifies a sensitivity loss.

If the background is \textit{overestimated} in the signal region the discovery potential is artificially reduced (lower two panels, solid lines). However, using the new method (dashed lines) it is raised and becomes closer to the sensitivity with the true model in hand. 
If one wishes to be more conservative, one can force $\epsilon$ to be positive, and the original distribution will remain almost untouched.
In case of \textit{underestimation}, the sensitivity is artificially enhanced.
Looking at the second and third panels in Fig.~\ref{fig:discovery_significance}, it can be seen that using the method of this paper (dashed line) the sensitivity is reduced to the level of the true model. This is due to the fact that the test statistic distribution is restored and can be described by the $\frac{1}{2}\chi^2$ distribution (Fig.~\ref{fig:discovery_user}). As expected, when applying this correction a loss of sensitivity is apparent and in the examples shown here is less than $10\%$. 

It can be seen that the proposed construction compensates for the unaccounted background shape without making any assumptions on its magnitude or functional shape.  
This is done naturally, without an inclusion of additional nuisance parameters.  
It is important to note that the distribution of the test statistic using the wrong model can not be estimated using MC as the true model out of which the data is really generated is unknown.

\subsection{Exclusion}\label{subsec:exclusion}
To illustrate the usage for an exclusion test we compute the 90\% Confidence Level (CL) upper limit distribution, using the true and false background models. In this case we present the results for two background variations: the most overestimated and the most underestimated ones.  
For each model we compute the sensitivity with and without the correction, and compare it with the sensitivity using the true model. 
We use the same setup as before with 100 events in the physics dataset, and 1000 of background calibration events.
As for discovery, the improper modeling will cause a breakdown of the $\frac{1}{2}\chi^2$ distribution of the test statistic,  leading to enhanced or reduced exclusion limits.

\begin{figure}[h!]
{\includegraphics[width=0.5\textwidth]{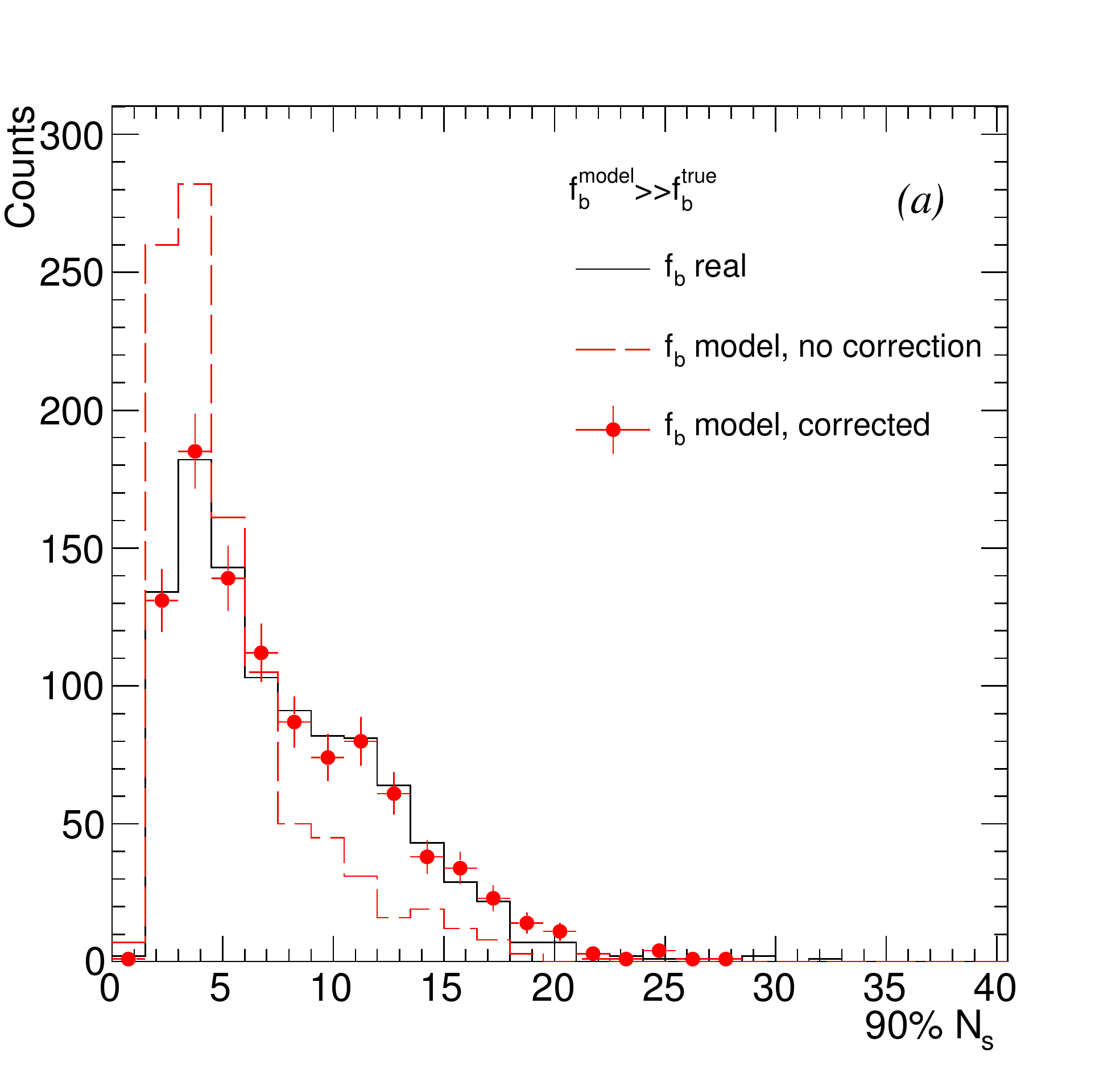}} 
{\includegraphics[width=0.5\textwidth]{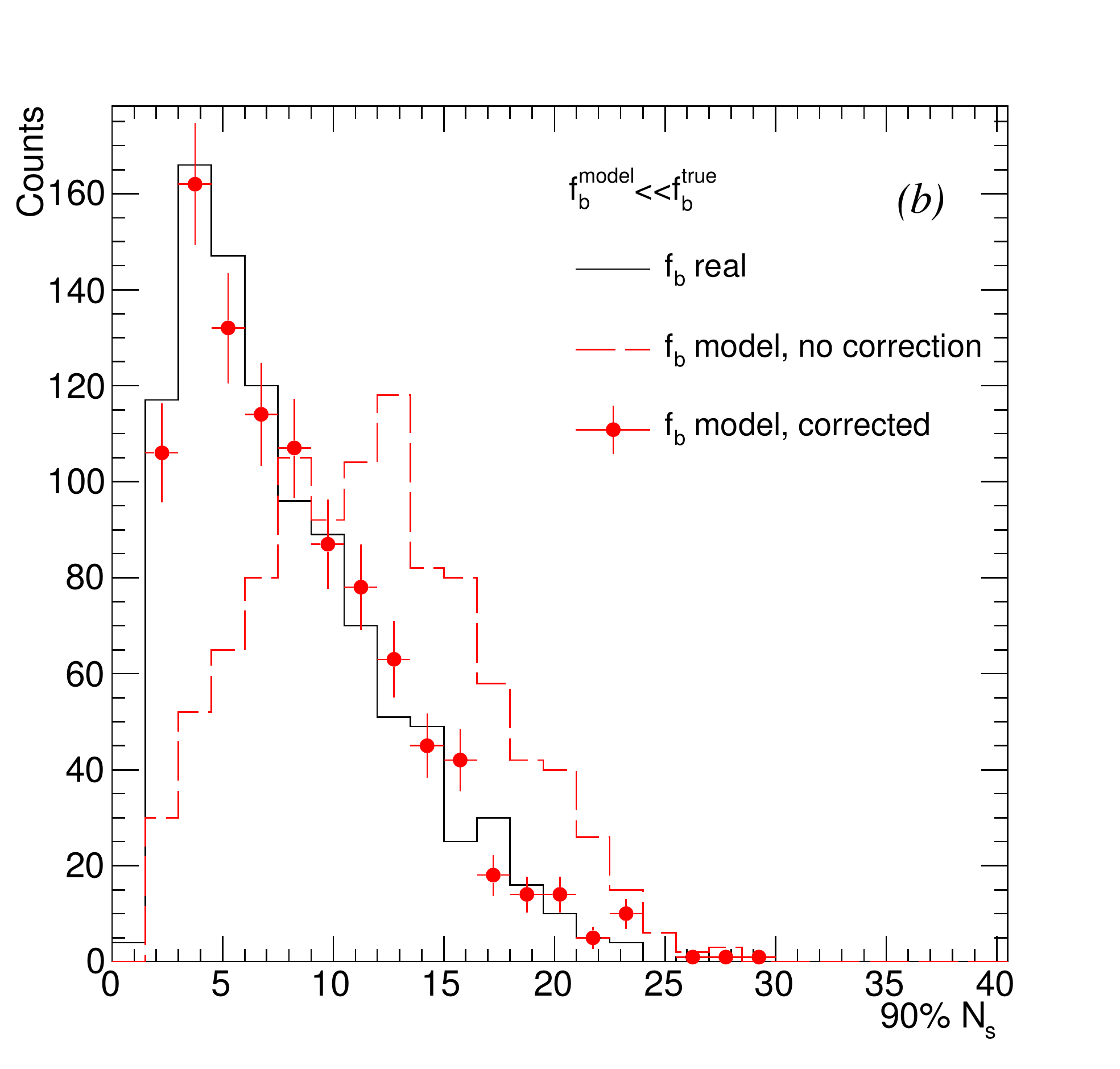}}
\caption{90\% CL exclusion limit comparison using the true background model (black), the wrong model (red dashed) and the wrong model plus correction (red circles). The construction presented in Sec.~\ref{sec:thesolution} prevents too good limits (left) in case of overestimated background, and prevents sensitivity loss in case of underestimated background (right).}
\label{fig:exclusion_user}
\end{figure}

Fig.~\ref{fig:exclusion_user} presents the tension between the limits using the wrong and the correct models. One can see that using the procedure defined in this paper cures the discrepancy.
If the model is overestimated  (Fig.~\ref{fig:exclusion_user} a), the sensitivity is artificially enhanced, and regions in the parameter space in which the experiment has no sensitivity might be excluded.
The correction restores the true sensitivity of the experiment.
Alternatively, if the background model is underestimated (Fig.~ \ref{fig:exclusion_user} b), the limits will be higher than the true one, and can be restored using the correction.

As before, to be more conservative, $\epsilon$ can be constrained to be negative throughout the procedure, and as a result the limits will remain practically unfixed in case of underestimation. For overestimation though, the limits will restore the true sensitivity.
 \section{Examples}\label{sec:examples}

In this section we present advanced examples in which the method defined in the previous section can be used. Many times, the functional shape of the background is defined by either a MC or by some analytic expression, and the parameters of the background are extracted using a signal-free calibration sample. In the first example, we show how the method accounts for the uncertainties on the functional shape, and fix the asymptotic behavior of the test statistic.

In the second example, the background is estimated in a nonparametric way, this is done in order to dodge wrong assumptions on the model. However, In case the calibration size is not large enough the asymptotic behavior will be broken, and can be restored using the method suggested in this paper.

The third example is a 2-dimensional case study. We show that the method works in higher dimensions as well, and we present a breakdown due to nonparametric estimation with small a calibration sample. 

As a general rule, all the applications depend on several aspects. 
A major aspect is the quality of the shape discrimination which can be quantified by $\alpha_d$ (see Eq.~\ref{eq:alphad} in appendix \ref{appendix:Atx}). We give this measure for each of the examples as reference.

\subsection{Functional Background model}
\label{subsec:examples_functional}
In many experiments the background model \fbm is constructed by fitting a function (analytic~\cite{Aprile:2012nq} or MC based~\cite{lux:2016}) to a control measurement (calibration sample) with one or more free parameters. This procedure assumes that the true background, \fbt, can be described by the chosen function and its parameters range.  In case the functional space spanned by \fbm does not cover \fbt, the results of the inference will be biased. In the following example we demonstrate how the method presented in this paper can resolve this issue. 

\begin{figure}[ht!]
{\includegraphics[width=0.33\textwidth]{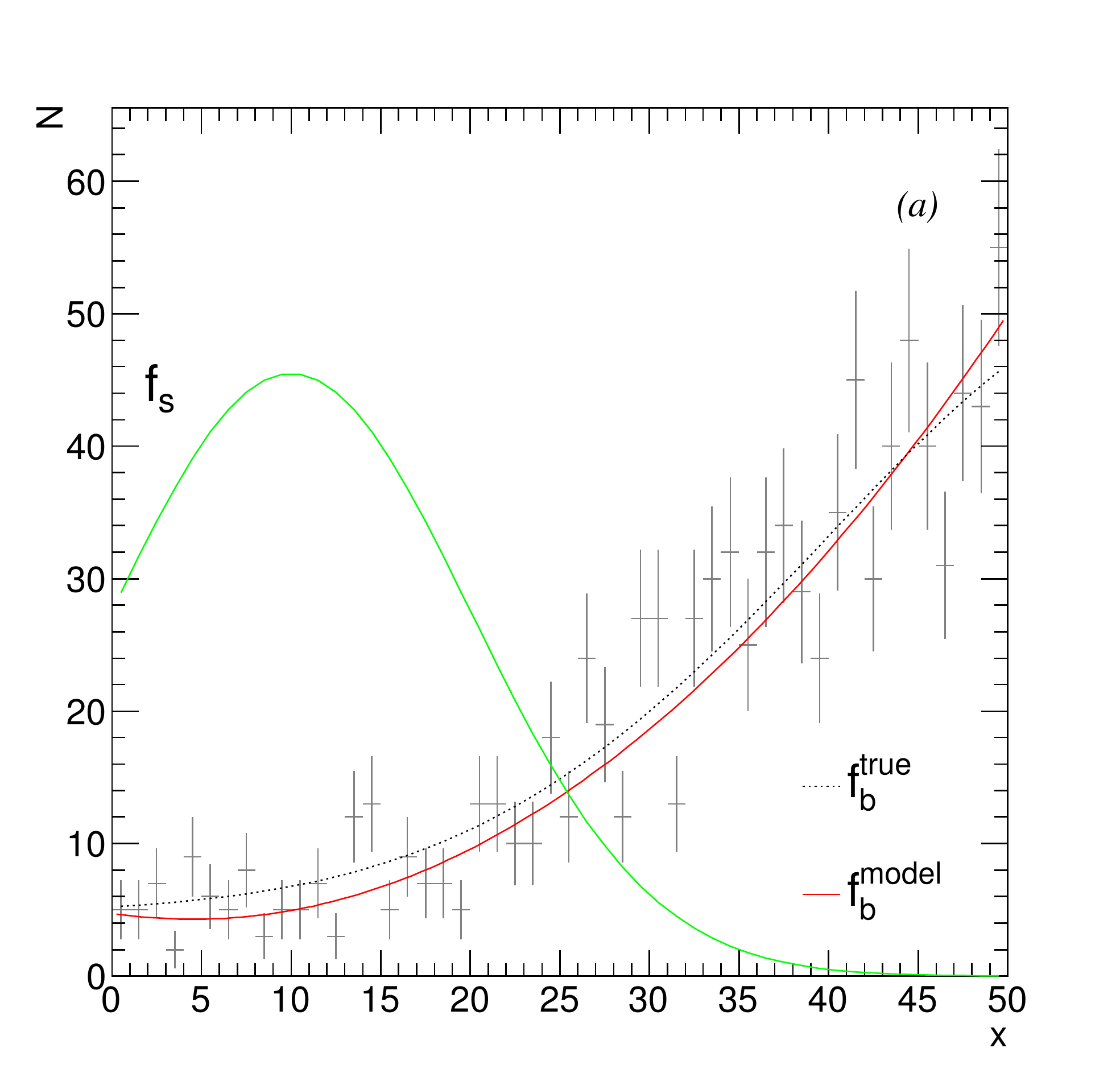}} 
{\includegraphics[width=0.33\textwidth]{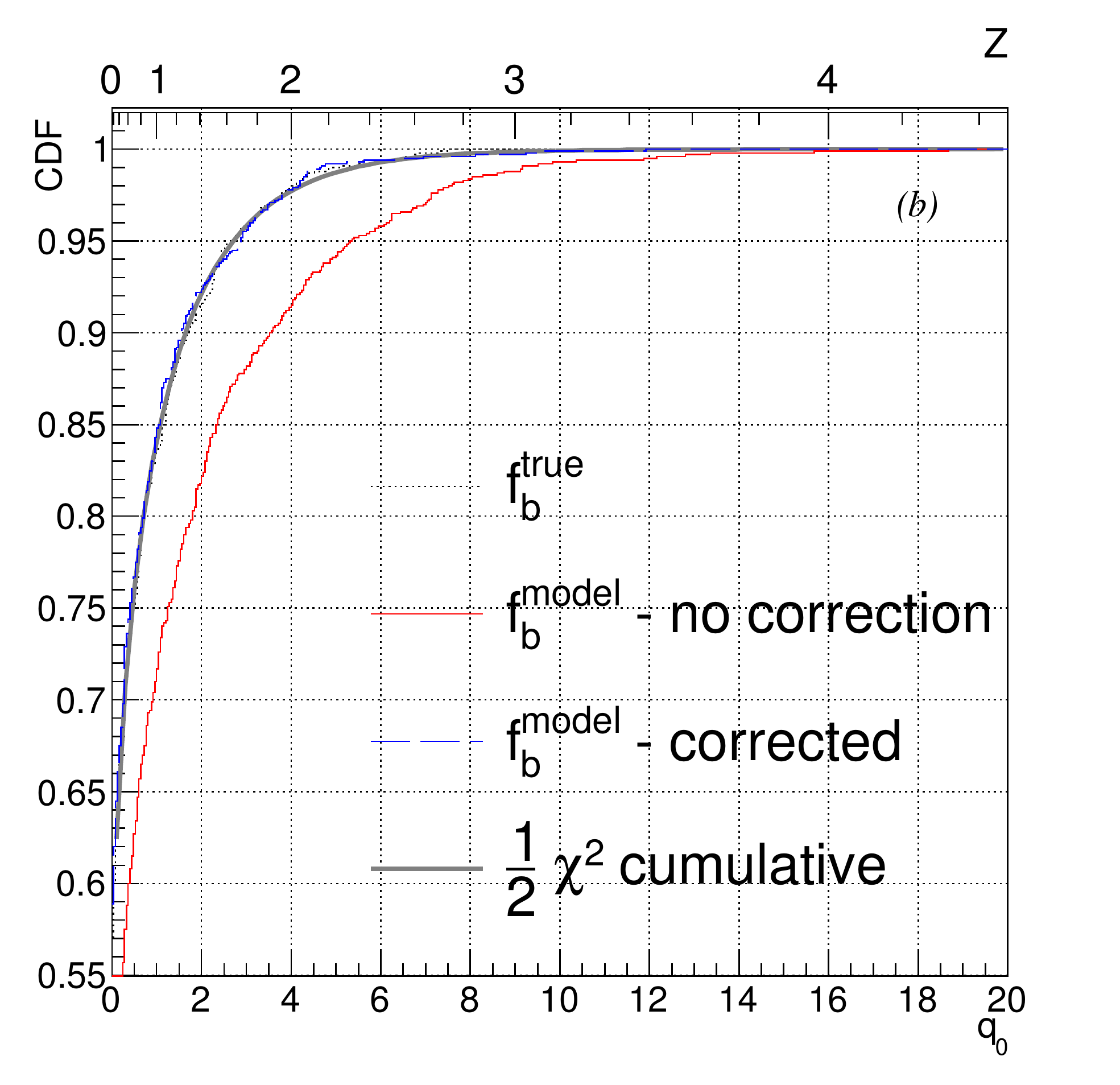}}
{\includegraphics[width=.33\linewidth]{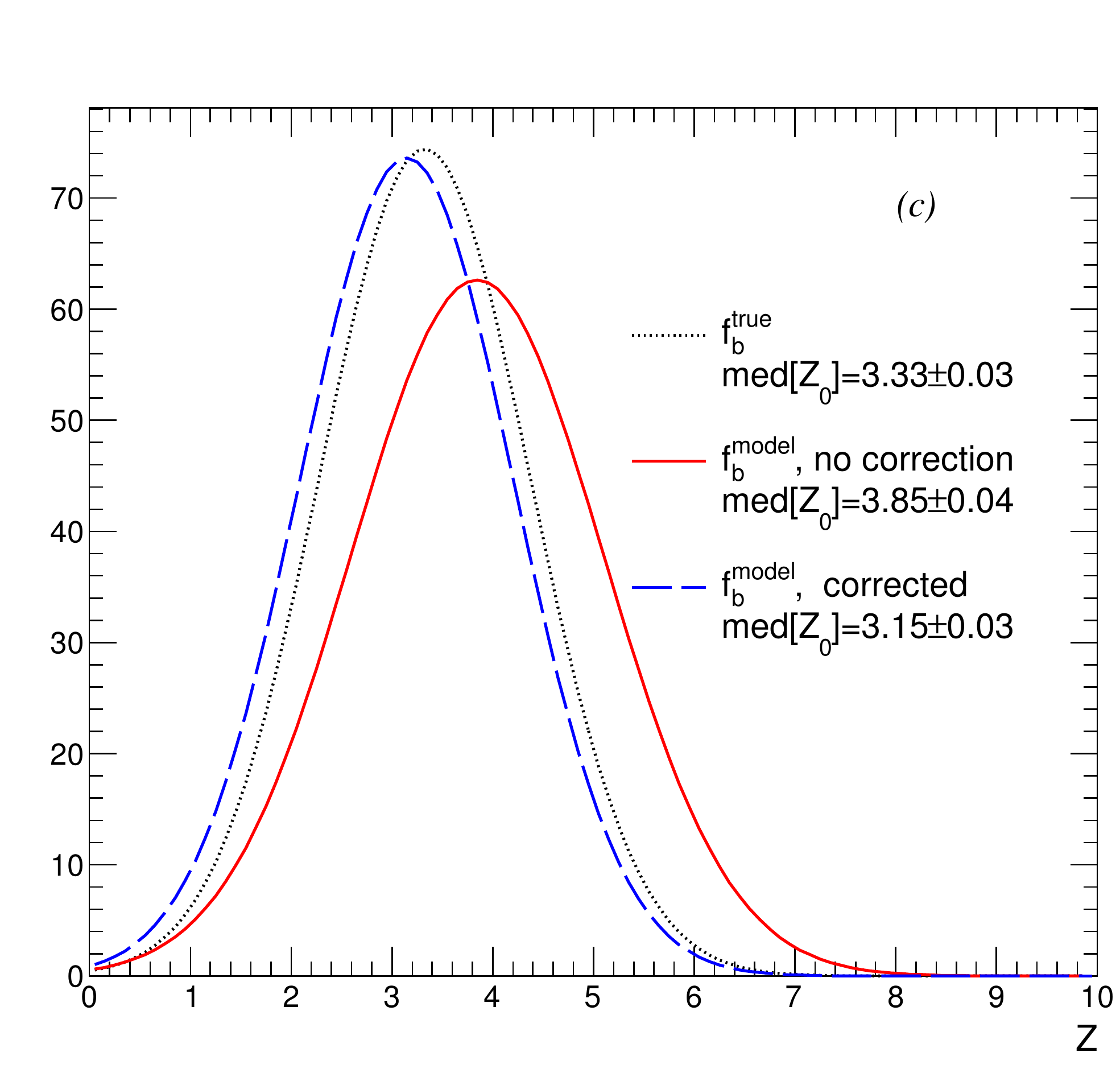}}
\caption{{\bf (a)}  One instance of a 1000 events calibration sample generated using \fbt (in black) and a the best fit to a different function \fbm (in red). The signal model is shown in green.  In this instance the background is underestimated in the signal region. {\bf (b)} Cumulative Distribution Function (CDF) of discovery significance for a signal-less dataset, $f(q_0|0)$. The expected $\frac{1}{2}\chi^2$  distribution is shown in gray, with the calculated distribution using \fbt (black). The red lines show the CDF using a best fit \fbm as described in the text, before applying the correction (red dashed) and after (blue solid). The undercoverage that causes enhanced false discovery rate is completely removed with the method. {\bf (c)} The discovery potential for 15 signal events. The distributions are presented for normal PL using the wrong model (solid red) and using the method presented in Sec.~\ref{sec:thesolution} (blue dashed). The distribution using \fbt is presented for comparison (gray dotted).}
\label{fig:functional}

\end{figure}

This is illustrated by constructing the background model using a fit to a calibration sample using different functional representations for \fbt and \fbm. Fig.~\ref{fig:functional} (a) shows an instance of a calibration dataset randomly generated from a tail of a Gaussian (\fbt), and its best fit to a second order polynomial (\fbm). 
It is noted that \fbm describes very well the background. In this example $\alpha_d=4.9$. 

We generate 1000 trials of calibration dataset (100 events) and physics dataset (100 events), where the physics dataset does not contain any signal.
For each dataset pair, the likelihood function was calculated using Eq.~\ref{eq:discoveryTestStat}. To enhance the effect we did not include the fit uncertainties in the likelihood function as nuisance parameters. This represents the case in which the model's parameter do not cover the true model.

Fig.~\ref{fig:functional} (b) shows the test statistic distribution and demonstrates the breakdown of the $\frac{1}{2}\chi^2$ distribution. In addition, it shows how the method of this paper corrects the undercoverage introduced by the faulty background model. Before applying the correction there is a clear undercoverage causing 4 times as many $2\sigma$ false discoveries than assumed, and around 2\%  of false $3\sigma$  discovery rate. 
After applying the correction the undercoverage is gone, and the $\frac{1}{2}\chi^2$ distribution is reproduced. 

For completeness, the discovery potential for 15 signal events over 100 background events is shown in Fig.~\ref{fig:functional} (c). It can be seen, by comparing the sensitivity to the one computed with \fbt, that the artificial enhancement of the sensitivity  is relaxed when using the method of this paper at the price of a small sensitivity loss. 

 \subsection{Empirical background model}\label{subsec:examples_empiricalModel}

There are cases in which a calibration sample is available and one wants to construct an empirical model based on this sample without any prior assumptions. The problem is that in those cases, there are no natural nuisance parameters that can model the uncertainties. In this section we show how the uncertainties can be modeled with the procedure described in this paper. In addition, we demonstrate how low calibration samples can lead to false discoveries if the procedure is not used. 

In case a calibration sample is available we can use the Kernel Density Estimation (KDE) in order to produce a nonparametric estimation of the background pdf. Let $\lbrace x_1,x_2,...,x_{N_{cal}}\rbrace$ be a calibration sample, the KDE is
\begin{equation}
\label{eq:KDE}
{f}^{KDE}_b(x) = \frac{1}{n} \sum_{i=1}^{n} K_h (x-x_i),
\end{equation}
where $K(x)$ is the \textit{kernel function}, a function that is normalized to one with a mean zero, and $h$ is a scale parameter called the \textit{bandwidth}. There are many different ways to choose the bandwidth while throughout this paper we choose to work with a variable bandwidth, inversely proportional to the events' density (a thorough introduction of KDE can be found e.g., in~\cite{Cranmer:2000du}).
The difficulty in using KDE for inference, and nonparametric estimators in general, is the lack of natural nuisance parameters. Then, any downward fluctuation of the calibration data in the signal region will result an enhanced probability for false discovery, due to the underestimated background model (similarly to subsection~\ref{subsec:userModel}). 

\begin{figure}[ht!]
{\includegraphics[width=0.5\textwidth]{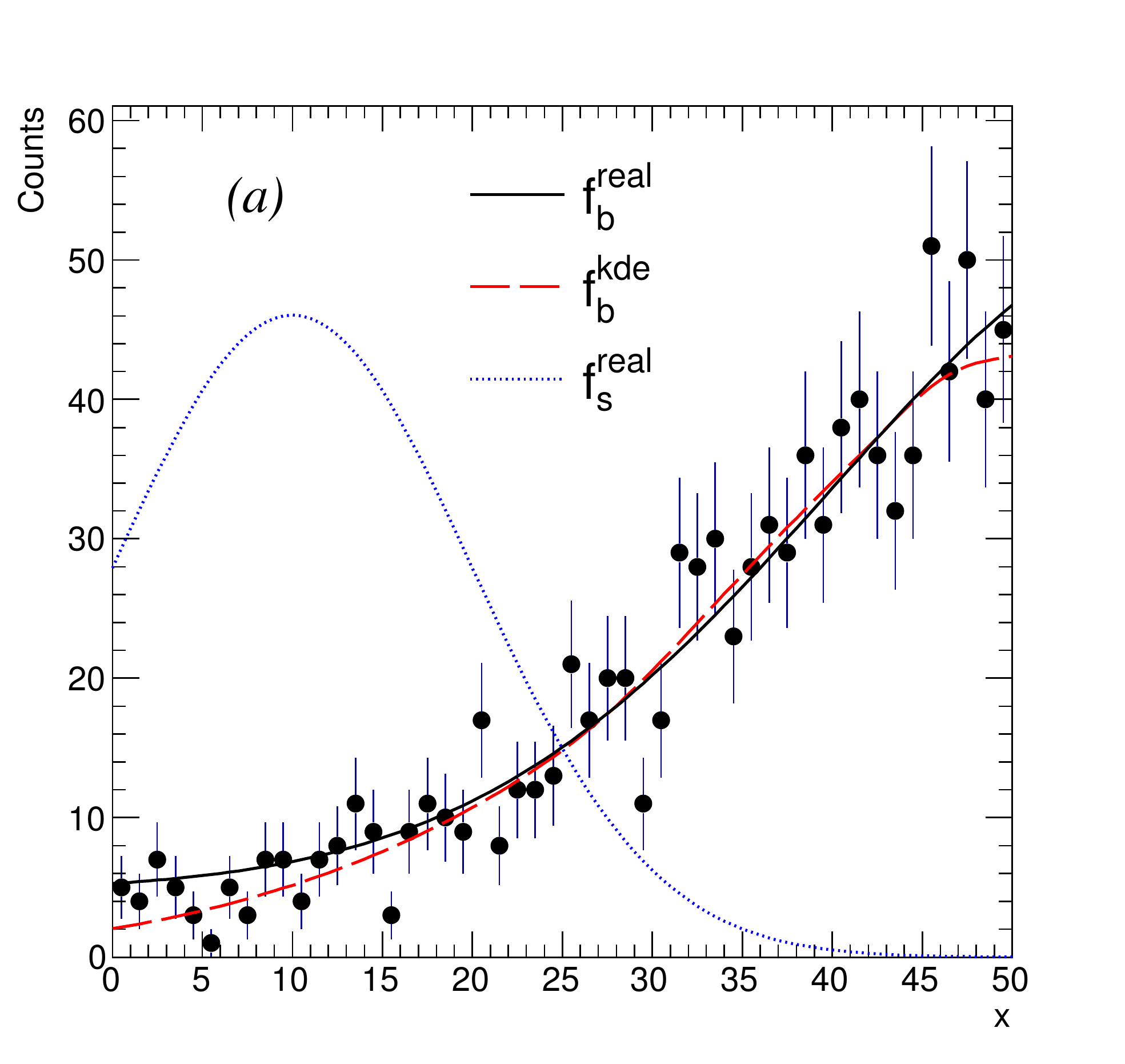}} 
{\includegraphics[width=0.5\textwidth]{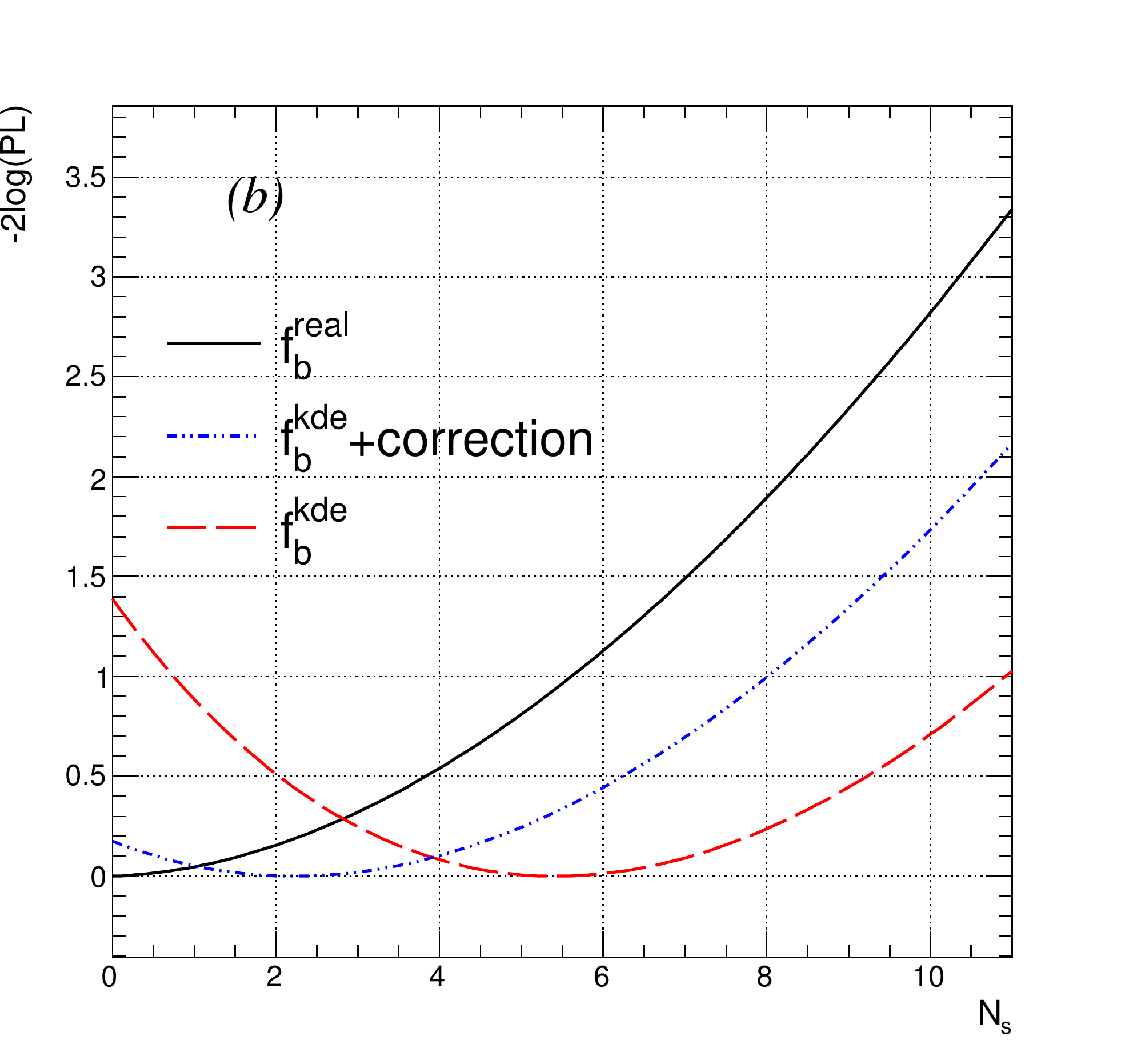}}
\caption{{\bf (a)}  One instance of KDE background model, $f^{KDE}_b$ (red dashed), constructed using 1000 calibration events. It can be compared with \fbt (solid). The signal distribution is also shown (blue dotted).  In this instance $f^{KDE}_b$ is underestimated in the signal region. {\bf (b)} The PL function using the the true background (black solid), and using the underestimated $f^{KDE}_b$ before applying the correction (red dashed). The dataset does not contain any signal events. The overestimation at $N_s=0$ is interpreted as a hint towards a signal like data, and is caused because of the underestimation of $f^{KDE}_b$ in the signal region. This can be avoided by the inclusion of the KDE uncertainties as explained in this section and illustrated by the corrected curve (dashed-dotted blue).}
\label{fig:kde_example_pl}
\end{figure}

An example of such instance is shown in Fig.~\ref{fig:kde_example_pl} (a), with $\alpha_d=4.9$.
Where a small downward fluctuation in the calibration dataset causes  an underestimation of the background in the signal region. This results in an overestimation of the profile likelihood function at $N_s=0$ as can be seen in Fig.~\ref{fig:kde_example_pl} (b).

\begin{figure}[ht!]
\begin{center}
{\includegraphics[width=0.45\textwidth]{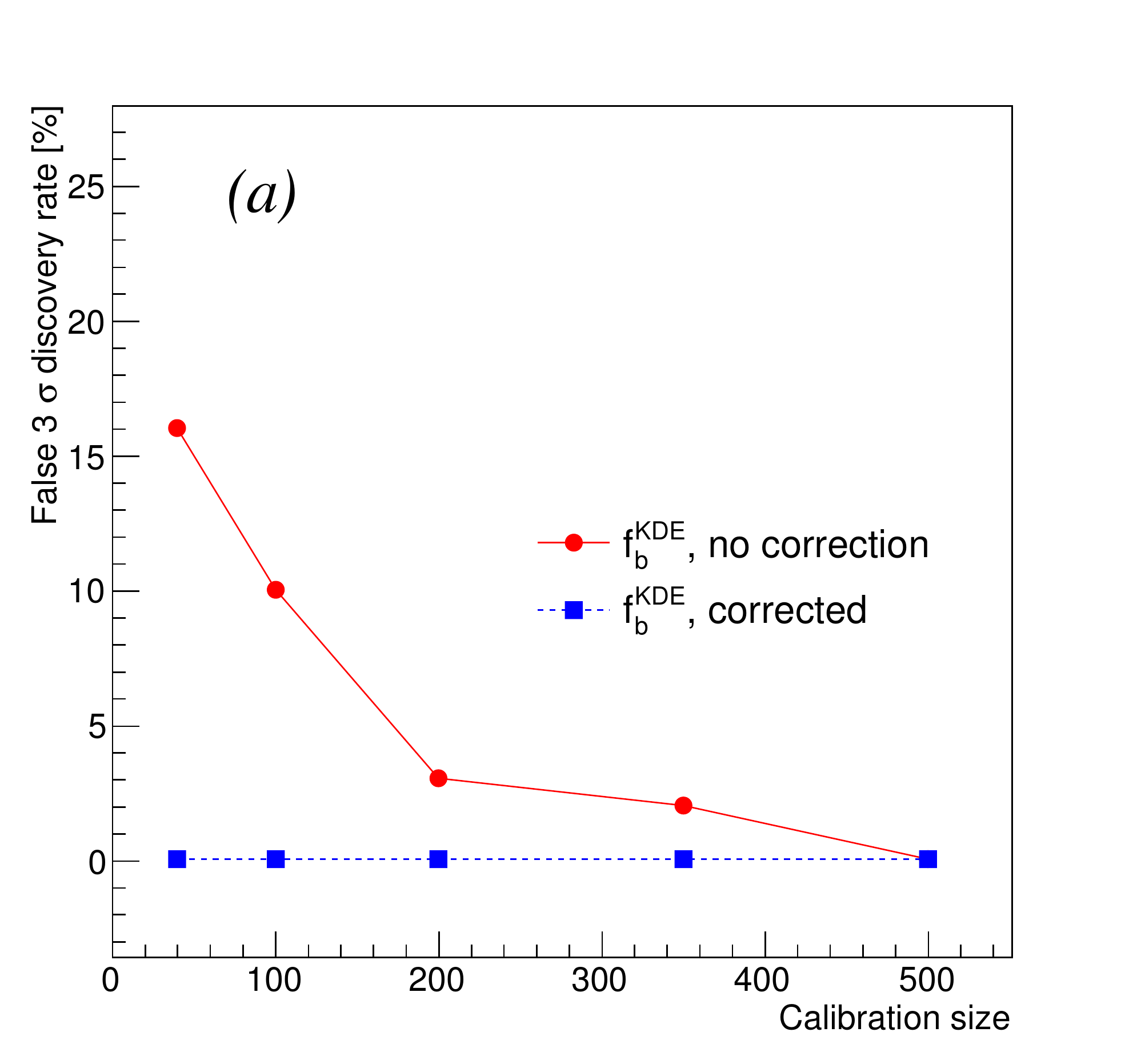}} 
{\includegraphics[width=0.45\textwidth]{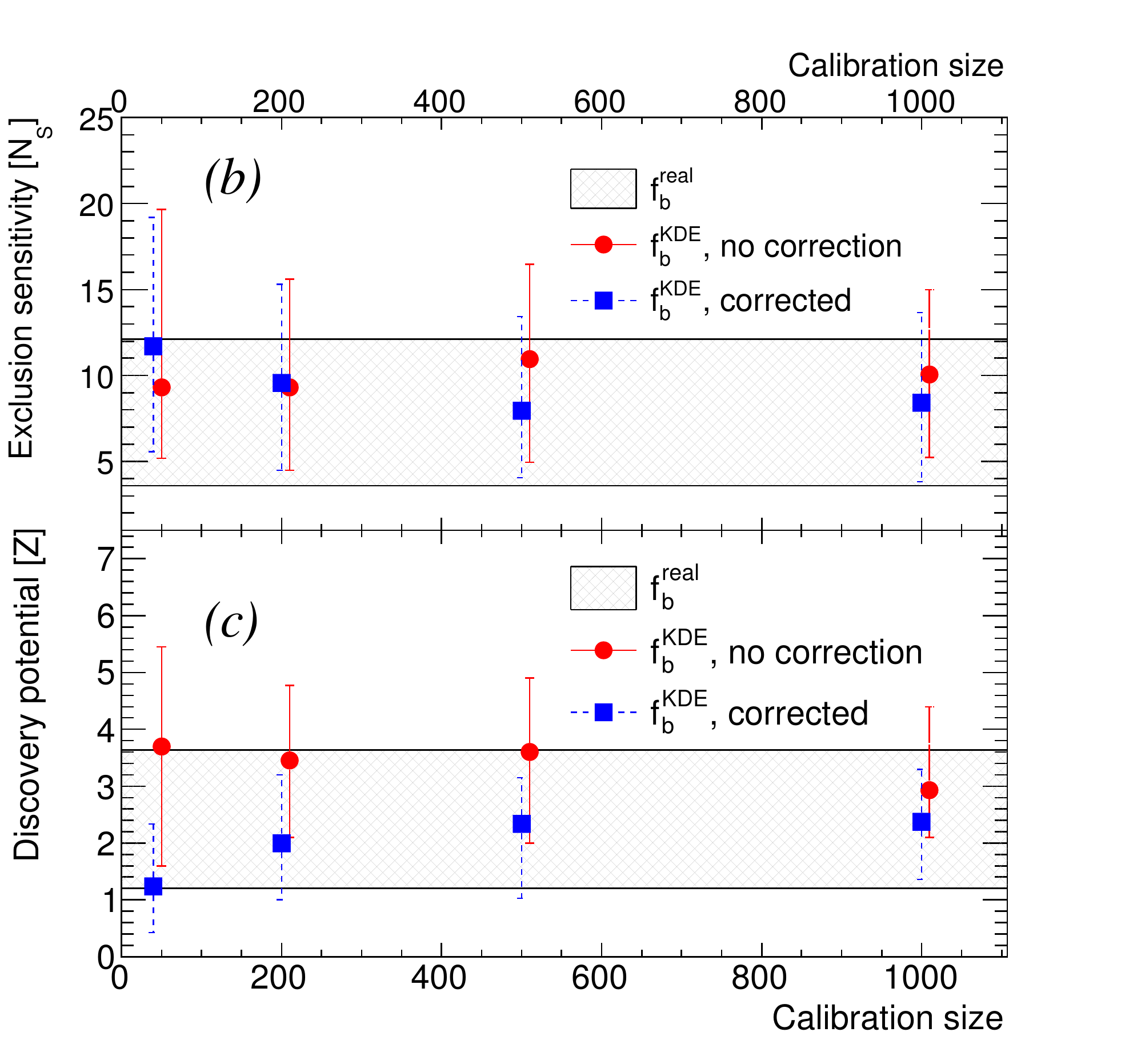}}
\end{center}
\caption{{\bf (a)}  False 3$\sigma$ discovery rate using KDE background model, $f^{KDE}_b$ (red line). The false discovery is removed using the method described in this paper as shown by the blue dashed line. {\bf (b)} 
 Exclusion sensitivity using $f^{KDE}_b$ (red circles), and using $f^{KDE}_b$ plus correction (blue squares). The median 90\% CL limit is shown along with its $\pm$1$\sigma$. The 1$\sigma$ sensitivity band using the true background model is shown for comparison (gray band). {\bf (c)} Discovery potential using PL with the true background (gray band), using $f^{KDE}_b$ (red circle), and using $f^{KDE}_b$ plus correction (blue squares). It is computed for 15 expected signal events over 100 expected background events.  }
\label{fig:kde_discovery_comparison}
\label{fig:KDE_exclusion_comparison}
\end{figure}

The false discovery presented in Fig.~\ref{fig:kde_example_pl} is caused by the breakdown of the $\frac{1}{2}\chi^2$ approximation. As for the previous example, this assumption on the test statistic distribution does not hold due to the improper model. This breakdown causes an enhanced false discovery rate. The 3$\sigma$ false discovery rate is presented in Fig.~\ref{fig:kde_discovery_comparison} (a) as a function of the calibration size. As the calibration size is increased the KDE approaches \fbt and the false discovery rate is reduced. However, using the method described in this paper the false discovery can be corrected also for insufficient calibration sample size.

Another important property of the procedure is its discovery potential. Ideally one would like to have a procedure that punishes lack of information by reduction in sensitivity. On the other hand, one does not wish to lose sensitivity  in case there exists enough knowledge. Our method is compared with the naive procedure, and with a PL using the \fbt in Fig.~
~\ref{fig:kde_discovery_comparison} (b). The discovery potential is shown for various calibration sizes. It is computed by the injection of 15 signal events over 100 background events.
For large calibration sample the method approaches the sensitivity of the true model and of the KDE. However, for small calibration samples the KDE artificially enhances the sensitivity, while the correction reduces it.
This is the penalty for lack of knowledge. 

Last but not least, we examine the behavior of the new method for setting exclusion limits. In Fig.~\ref{fig:KDE_exclusion_comparison} we compare the 90\% CL exclusion limit of the method with PL using KDE (red squares), and with PL using the true model (gray band). It can be seen that the sensitivity is comparable to KDE for small calibration sizes, however it approaches the sensitivity of the true model for large calibration sizes.

 \subsection{Multi-Dimensional empirical background model}
\label{subsec:examples_2d}
A third example is motivated by experiments trying to directly detect dark matter \cite{Undagoitia:2015gya}. These experiments typically expect signal in the order of a few events per year or less, and determining the background PDF lacks in general  a complete physically motivated parametric model. Therefore the PDF is, in many cases, estimated by a dedicated calibration dataset, see e.g., \cite{Aprile:2012nq}. In other cases a model driven background model is used, as in \cite{lux:2016}, where a complementary 5 dimensional unbinned analysis is performed.
 
\begin{figure}[ht!]
\centerline{\includegraphics[width=0.5\textwidth]{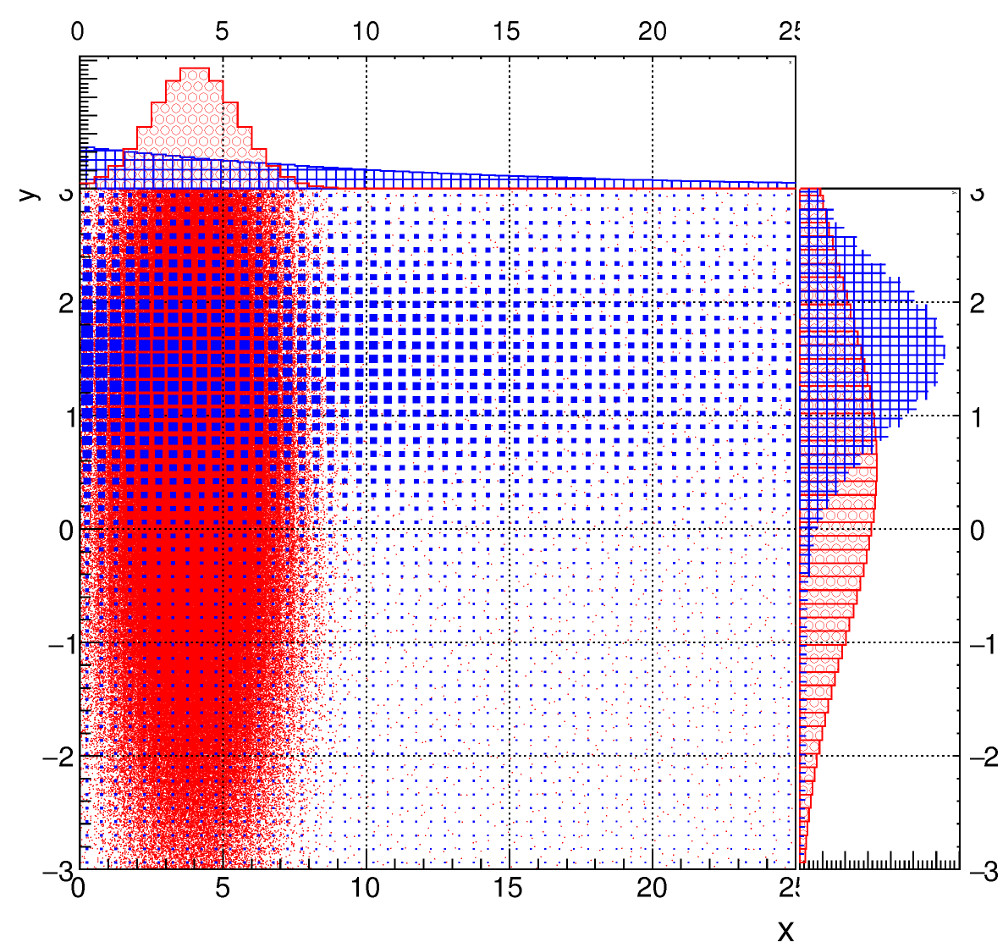}}
\caption{2D example, the true PDFs of signal (red) and background (blue), with projection on the $x$ and $y$ axes. The two dimensional discrimination between the two is based on the combination of them. }
\label{fig:2d}
\end{figure}

For this example\footnote{For this section the RooFit toolkit was used~\cite{Verkerke:2003ir}} we present the true PDFs of a two dimensional discrimination space in Fig.~\ref{fig:2d}. In the case presented here $\alpha_d=7$. The two dimensions can represent two measured channels (e.g., charge and light, or heat and charge) or a combination of them. The natural invariance under transformations of the unbinned PL makes the specific choice of axes irrelevant.

We use a two dimensional adaptive KDE procedure to estimate the background model, similarly to the previous example. For each instance, we draw a number of calibration events and infer a background model, and then use this model for statistical inference of a 100 background events, with and without added signal of 15 events, both numbers subject to Poisson fluctuations. 

\begin{figure}[ht!]
\includegraphics[width=0.48\textwidth]{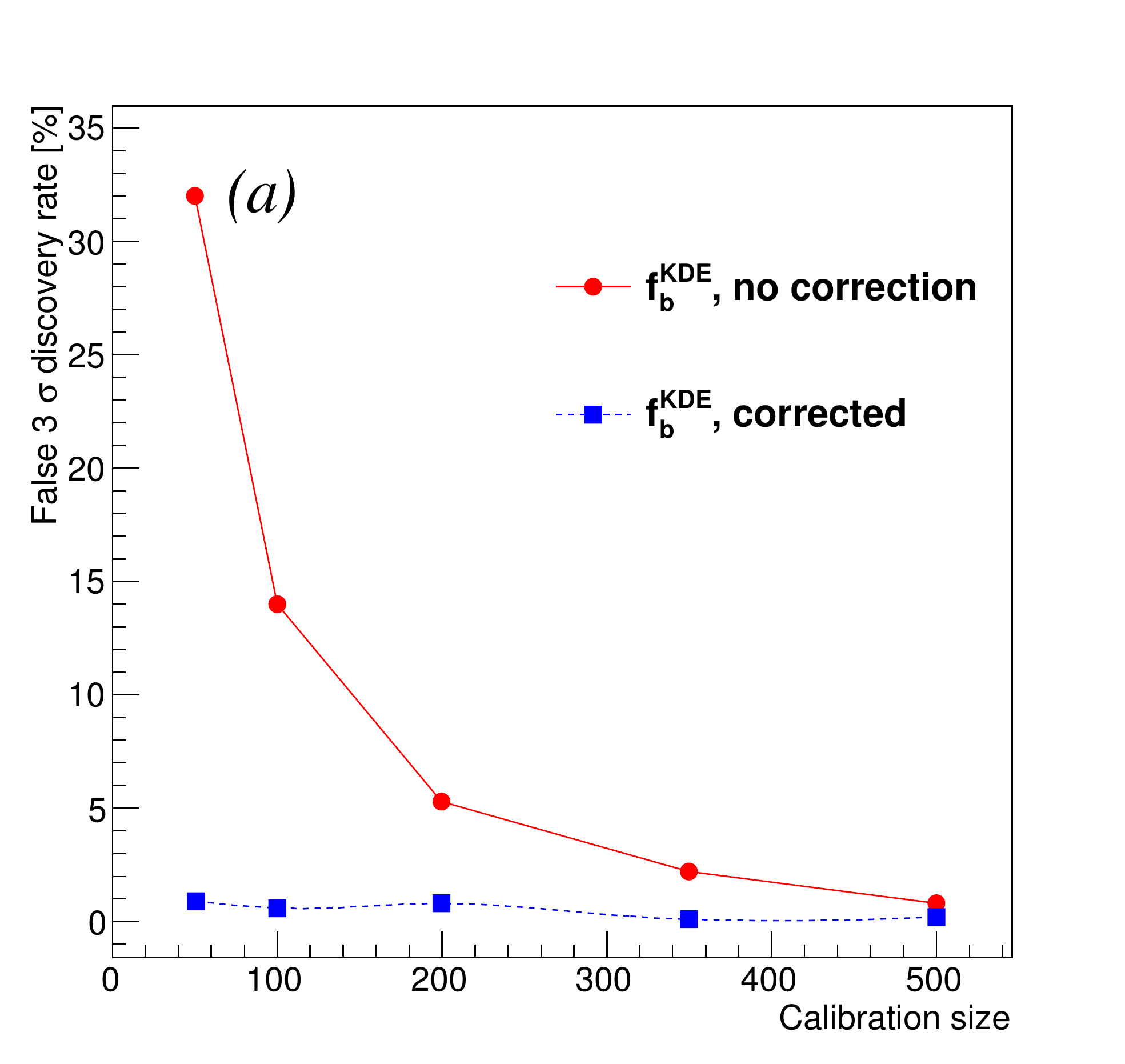}
\includegraphics[width=0.48\textwidth]{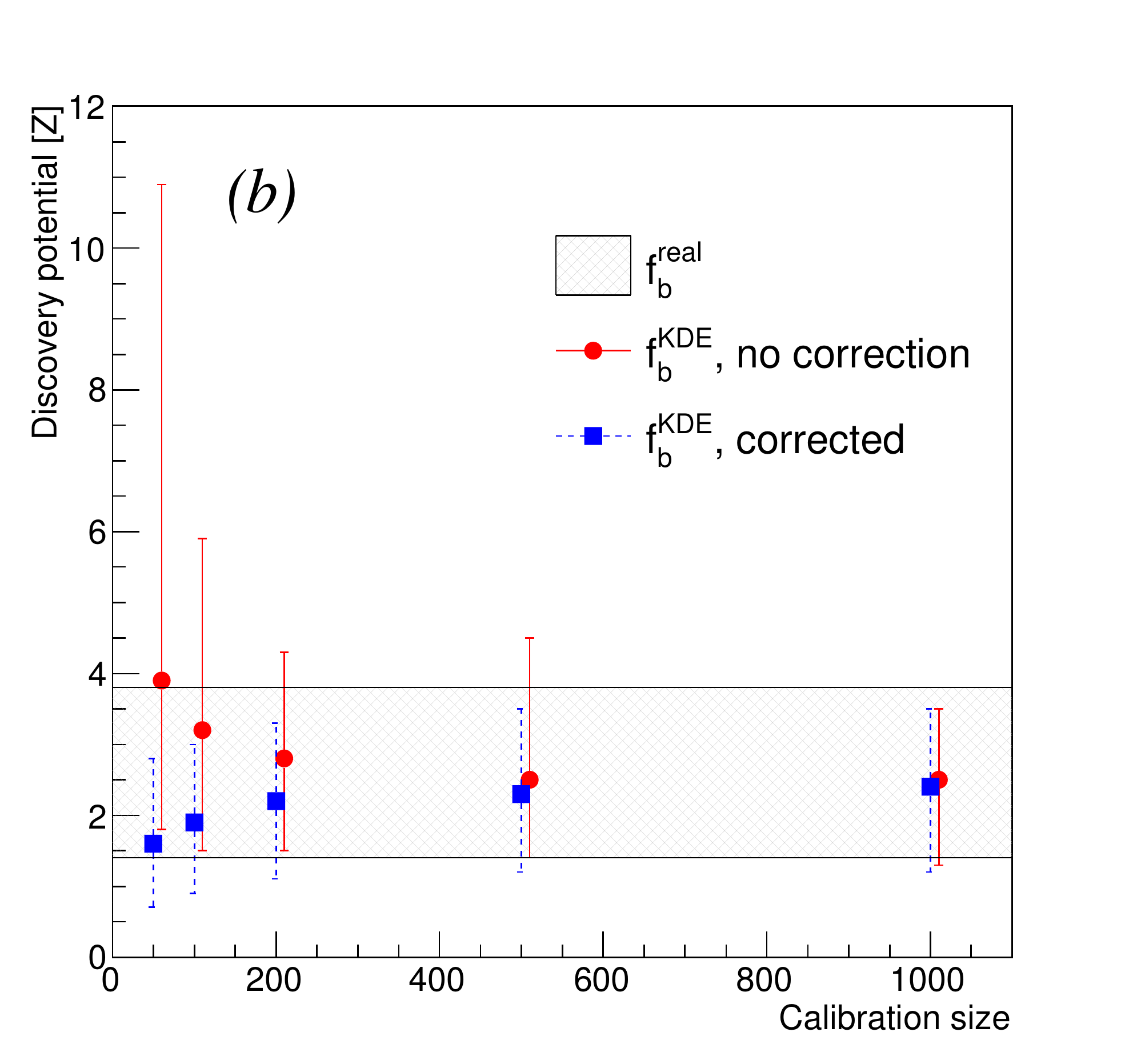}
\caption{2D example results. {\bf (a)}  False 3$\sigma$ discovery rate using KDE background model, $f^{KDE}_b$ (red line). The false discovery is removed using the method described in this paper as shown by the blue dashed line. {\bf (b)} 
Discovery potential using PL with the true background (gray band), using $f^{KDE}_b$ (red circles), and using $f^{KDE}_b$ plus correction (blue squares). It is computed for 15 expected signal events over 100 expected background events. }
\label{fig:2dresults}
\end{figure}

Fig.~\ref{fig:2dresults} summarizes the results of this study. On the left panel one can see the case of no signal, and how often a $3\sigma$ false detection may occur with and without the correction presented in this paper, as a function of the calibration size. Comparing this to the one dimensional example, it is pointing to a larger calibration sample needed when increasing dimensionality - even though the discrimination is somewhat better in the 2D case ($\alpha_d$ of 7 vs. 4.9). The right panel shows the discovery potential of an injected signal, which for low calibration statistics exceeds the expected one by the true PDF. Again, the discrepancy for two dimensions is larger than in one dimension for a given calibration size. In both cases we see that the corrected PL is curing the false or overly confident discovery, but reduces accordingly the sensitivity - a natural result of less knowledge regarding the true nature of the background.

\section{Summary}
\label{sec:summary}
In this paper we have presented a new method that protects against biases created by mismodeling of the background.
 We show that statistical inference without this addition is prone to risks of false discovery and artificially enhanced sensitivity, which are cured when using Eqs.~\ref{eq:correctedLikelihood} and~\ref{eq:EpsilonLikelihood}. We work within the PL framework and present the asymptotic behavior restoration and explore a number of examples, using parametric models, non-parametric models and two dimensional  parameter space. Additional complexity and dimensionality of the data would increase the dangers of mismodeling, and make this addition an essential part of unbinned PL usage. Application of the method to more complicated scenarios such as external data coming from other experiments can be done with proper care. One must bear in mind that uncertainties attached to detector performance or yet-unknown physics effects can not be eliminated by a model or simulation, however numerically accurate it is. 

\section*{Acknowledgments}
\label{sec:acknowledgements}
The authors would like to thank Jan Conrad, Eilam Gross and Daniel Lellouch for fruitful discussions.  We are also grateful to Teresa Marrodán Undagoitia and Manfred Lindner for their scientific support. This work was supported by ISF I-CORE grant 1937/12 ``The quantum Universe'' and the DFG research training group ``Particle physics beyond the standard model''.

 \clearpage

\appendix

\section*{Appendix}
\section{The interplay between binned and unbinned likelihood}\label{appendix:A}

\subsection{The log likelihood function}
\label{appendix:Abinned}
Here we will demonstrate the similarity between the binned and the unbinned likelihood function.

Starting with the binned-likelihood function defined in Eq.~\ref{eq:bLikelihood} we can obtain the log-likelihood function

\begin{align*}
\label{eq:bLikelihood}
\mathrm{log}(\Like_b) &= \mathrm{log}\left[\prod_j^{N_{bins}} \mathrm{Poiss}(n_j|\epsilon_s^j N_s + \epsilon_b^j N_b) \right]=\\
& \quad \sum_j^{N_{bins}} \left[ -\epsilon_s^j N_s - \epsilon_b^j N_b +  n_j \mathrm{log}(N_s \epsilon^j_{s} + N_b \epsilon^j_{b})\right]=\\
& \quad - N_s - N_b + \sum_j^{N_{bins}} n_j \mathrm{log}(N_s \epsilon^j_{s} + N_b \epsilon^j_{b}),
\end{align*}
where we use the fact that $\sum_{j}\epsilon_{j}=1$, and neglect constant terms.

Alternatively, starting from Eq.~\ref{eq:ubLikelihood}, 
the unbinned log-likelihood function can be projected into bins. This is done by the replacement of $f(x_i)$ with a binned probability density function, namely it can be replaced by $\epsilon^{bin(x_i)}/W$, the probability to be found in the bin, divided by the bin width, $W$.

\begin{align*}
\mathrm{log}(\Like_{ub}) &=   \mathrm{\mathrm{log}}\left[ \mathrm{Poiss}(N|N_s+N_b)\prod_i^N \frac{N_s f_s(x_i) + N_b f_b(x_i)}{N_s+N_b}\right] = \\
& \quad -N_s-N_b+\sum_{i=1}^N \mathrm{log}(N_s f_s(x_i) + N_b f_b(x_i))= \\
& \quad -N_s -N_b +\sum_{i=1}^N \mathrm{log}(N_s \epsilon^{bin(x_i)}_{s}/W + N_b \epsilon^{bin(x_i)}_{b}/W) = \\ 
& \quad - N_s - N_b + \sum_j^{bins} \sum_{i=1}^{n_j} \mathrm{log}(N_s\epsilon^j_{s} + N_b \epsilon^j_{b}) + constant=\\
& \quad - N_s - N_b + \sum_j^{N_{bins}} n_j \mathrm{log}(N_s \epsilon^j_{s} + N_b \epsilon^j_{b}) + constant,
\end{align*}

which yields an identical expression to the binned case, up to a constant.

 \subsection{Discrimination space}
\label{appendix:Atx}
In general, it is quite common to measure more than one variable for each event. Out of those variables one would usually like to get the maximum discrimination between signal and background. Namely, to have the best sensitivity on the parameter of interest, $N_s$.
This question can be phrased in the following way: assuming that we have some data space $\vec{x}$ in which the measurement has been done. We can apply transformation on the variables and work in other space $\vec{u}$. What is the space in which we get the smallest $\sigma_{N_s}$.

Working with Eq.~\ref{eq:ubLikelihood}, $\sigma_{Ns}$ can be estimated by calculating the expectation values of the second derivatives of the log-likelihood function
\begin{equation}
 \label{eq:sensitivityNsNb}
  \begin{gathered}
 \alpha \equiv -E\Big{[}\frac{\partial^2 {\LL}}{\partial {N_s}^2}\Big{]},\\
 \beta \equiv -E\Big{[}\frac{\partial^2 {\LL}}{\partial N_s \partial N_b}\Big{]},\\
 \gamma \equiv  -E\Big{[}\frac{\partial^2 {\LL}}{\partial {N_b}^2}\Big{]}.
  \end{gathered}
 \end{equation}

 The resolution of $N_s$ can be computed to be $\sigma_{N_s}^2=1/(\alpha-\beta^2/\gamma)$. 

For the unbinned likelihood it can be easily shown that $\alpha$, $\beta$ and $\gamma$ are invariant under coordinate transformation - hence we cannot gain or loose sensitivity by choice of coordinates

\begin{equation}
 \label{eq:sensitivityNsNb2}
  \begin{gathered}
 -E\Big{[}\frac{\partial^2 {\LL}}{\partial {N_s}^2}\Big{]} = \int \frac{f_s^2(x) }{N_s f_s(x) + N_b f_b(x)}dx \equiv \alpha,\\
 -E\Big{[}\frac{\partial^2 {\LL}}{\partial N_s \partial N_b}\Big{]} = \int \frac{f_s(x) f_b(x)}{N_s f_s(x) + N_b f_b(x)}dx \equiv \beta,\\
 -E\Big{[}\frac{\partial^2 {\LL}}{\partial {N_b}^2}\Big{]} = \int \frac{f_b^2(x) }{N_s f_s(x) + N_b f_b(x)}dx \equiv \gamma.
  \end{gathered}
 \end{equation}
 
The resolution on $N_s$ can be computed to be $\sigma_{N_s}^2=1/(\alpha-\beta^2/\gamma)$, and we see that $\alpha$, $\beta$ and $\gamma$ are invariant under transformation of $\vec{x}$. This means that for an unbinned analysis the sensitivity,  $\sigma_{N_s}$, is optimal and invariant in any space we work in. This is another advantage of this procedure.

When rejecting the null hypothesis with Eqs.~\ref{eq:sensitivityNsNb2}, $\beta=\gamma=N_b^{-1}$ and we are left with the estimator for {\it shape discrimination}
\begin{equation}\label{eq:alphad}
\alpha_d\equiv\int\frac{f_s^2(x)}{f_b(x)}dx,
\end{equation}
ranging from 1 to infinity, where $\alpha_d=1$ if and only if $f_s=f_b$. The higher $\alpha_d$, the better the shape discrimination between the background and the signal. This is again an invariant under transformations of the measured space. In the limit $\alpha_d\rightarrow\infty$, PL does not contribute to the sensitivity beyond simple counting, as the background leakage goes to zero.

\section{Motivation for $\epsilon$ as a nuisance parameter}\label{appendix:B}
\subsection{Extended likelihood on the calibration}\label{appendix:B1}

Lets assume that we have a control measurement consists of $n_c$ background events. The unbinned likelihood (Eq.~\ref{eq:ubLikelihood}) can be written in order to constrain the background model. The calibration likelihood function is 
\begin{equation}
\label{eq:BcalLikelihood}
 \Like_{cal} = \mathrm{Poiss}(n_{c} | n_s + n_b) \prod_{i=1}^{n_{c}} \frac{n_s  f_s(x_i) +  n_b f_b(x_i) }{n_s + n_b} ,
\end{equation}
where $n_s$ and $n_b$ are the expected number of signal and background events.
The $n_s$ component is introduced in order to quantify the level of signal like behavior of the background model, as we don't expect to have any signal events in the calibration dataset.
In the case of good background over signal discrimination $n_s\rightarrow 0$ as $n_c$ increases. 

The log likelihood, after neglecting constant offsets, is
\begin{equation}
\label{eq:UBcalLikelihood}
\begin{split}
\mathrm{log}(\Like_{cal}) = & \sum_{i=1}^{n_{c}} \mathrm{log}(n_s f_s (x_i)  + n_b f_b(x_i) ) -n_s -n_b\\
=& \sum_{i=1}^{n_{c}}\mathrm{log}\Big( \epsilon f_s(x_i) + (1-\epsilon) f_b(x_i) \Big) -n + n_{cal} \cdot \mathrm{log}(n),
\end{split}
\end{equation}
with $\epsilon=n_s/(n_s+n_b)$ , and $n=n_s+n_b$.
We identify the term in the logarithm to be the background model. Namely instead of keeping $f_b(x)$ fixed we let it be contaminated with a signal like component, where $\epsilon$ is the contamination parameter, and the calibration dataset pose a constraint on it.

For a `physics' dataset, which might contain a signal we will use Eq.~\ref{eq:ubLikelihood}, with one change. We will propagate the background model as follows $$f_b(x) \rightarrow (1-\epsilon)f_b(x) + \epsilon f_s(x)$$, in order to reflect the uncertainty of the model. 
 
The log likelihood function for the `physics` dataset is 
\begin{equation}
\mathrm{log}(\Like_{phys}) = \sum_{i=1}^{N} \mathrm{log}\Big( (N_s+\epsilon N_b) f_s (x_i) + (N_b-\epsilon N_b) f_b(x_i) \Big) - N_s - N_b,
\end{equation}
and the combined log likelihood becomes
\begin{equation}
\label{eq:newLikelihood} 
\begin{split}
\mathrm{log}(\Like_{cal}\times\Like_{phys}) & = \sum_{i=1}^{N} \mathrm{log}\Big( f_s (x_i) (N_s + \epsilon N_b) + f_b(x_i) (N_b-\epsilon N_b) \Big) + \\ 
& \sum_{i=1}^{n_{c}}\mathrm{log}\Big(\epsilon f_s(x_i) +(1-\epsilon)f_b(x_i)\Big) -N_s - N_b  -n + n_{c}\mathrm{log}(n),
\end{split}
\end{equation}

where $N_s$ is the parameter of interest, and  $n$, $\epsilon$, and $N_b$ are nuisance parameters. 
$n$ is constrained to be $n=n_{cal}$ independently of the other variables, hence the last two terms can be neglected.

\subsection{Likelihood functional}\label{appendix:B2}
Lets take the unbinned likelihood function $\Like(N_s,N_b)$ (Eq.~\ref{eq:ubLikelihood}) and treat the background model itself as a nuisance parameter.
Now, we have a semiparametric likelihood functional $\Like(N_s,N_b,f_b(x))$, with two parameters, $N_s$ and $N_b$, and one nuisance function, $f_b(x)$, where the parameter of interest is $N_s$.

For simplicity we assume that $N_b$ is known, and check the 
first order variation with respect to $N_s$ and $f_b$ in order to identify functional variation on the background model that can be used.

\begin{equation}
  \begin{gathered}
 \frac{\partial \mathrm{log}(\Like)}{\partial N_s} \Delta N_s = - \Delta N_s +\sum_i \frac{ \Delta N_s}{N_s f_s(x_i) + N_b f_b(x_i)} f_s(x_i) ,\\
 \frac{\delta \mathrm{log}(\Like)}{\delta N_s} \phi(x) = \sum_i \frac{N_b}{N_s f_s(x_i) + N_b f_b(x_i)} \phi(x_i).
  \end{gathered}
\end{equation}

For discovery we take a negative variation of $N_s$ and we are scanning below $\hat{N_s}$. Therefore the absolute value of the second term is greater than the value of the first term. Then, if we allow variation of the background model, by setting $\phi(x) \propto f_s(x)$ we relax the difference between the two terms for any given instance of the data $\lbrace x_1,x_2,...,x_N \rbrace$.

 \clearpage

\end{document}